\documentclass[%
 reprint,
 amsmath,amssymb,
 aps,
showkeys,
]{revtex4-2}

\usepackage{graphicx}
\usepackage{dcolumn}
\usepackage{bm}
\usepackage{hyperref}
\usepackage{amsmath}
\usepackage{color}
\usepackage[dvipsnames]{xcolor}
\usepackage[normalem]{ulem}

 \definecolor{AMK}{RGB}{220, 10, 10}

\begin{document}

\preprint{}

\title{Spatial asymmetry of optically excited spin waves in anisotropic ferromagnetic film}

\author{N. E. Khokhlov}
 \email{n.e.khokhlov@mail.ioffe.ru}
 \homepage{https://ioffe.ru/ferrolab/}
\author{Ia. A. Filatov}
\author{A. M. Kalashnikova}

\affiliation{%
    Ioffe Institite, 26 Politekhnicheskaya, 194021, St. Petersburg, Russia
}%

\date{\today}

\begin{abstract}
We analytically discuss and micromagnetically prove the ways to tune the spatial asymmetry of the initial phase, amplitude, and wavevectors of magnetostatic waves driven by ultrafast laser excitation. 
We consider that the optical pulse heats a thin ferromagnetic metallic film and abruptly decreases the saturation magnetization and the parameter of uniaxial anisotropy.
The two corresponding terms of laser-induced torque have different azimuthal symmetries, with the 4-fold symmetry of the demagnetization-related term, and the isotropic distribution of the anisotropy-related term.
As a result, the initial phase and amplitude of excited magnetostatic waves have a non-trivial azimuthal distribution tunable with the angle between the external magnetic field and anisotropy axis, and the laser spot diameter.
Moreover, the variation of these parameters tunes the distribution of wavevectors, resulting in additional asymmetry between the spectral components of the waves propagating in different directions.
\end{abstract}


\maketitle


\section*{\label{sec:intro} Introduction}

Magnonics is an emerging accompaniment to electronics, where spin waves (SW), $i.e.$ collective oscillations of spins, transfer information across magnetic nanostructures without charge currents \cite{Kruglyak_magnonicsJPhysD2010}.
This perspective field promises to solve a series of problems inherent for traditional electronics, such as high Joule energy losses, while bringing extra degrees of freedom characteristic for, $e.g.$ photonics.
Practical interest in magnonics is mainly due to its applications in telecommunications, since characteristic frequencies of SW lie in the range from sub-GHz to tens of THz.
Here, magnonics has an obvious advantage, since typical wavelengths of SW are several orders of magnitude smaller than the ones for electromagnetic waves of similar frequencies.
Moreover, the SW possess an extended flexibility of tuning their characteristics through the variation of external magnetic field, material properties, waveguide geometry, and even external stimuli \cite{chumak2022roadmap, grundler2015reconfigurable, Grachev_ReconfigurablePRAppl2023, Gubanov_FrequencyPRB2023}

To achieve an efficient control over SW, it is necessary to identify means to actively tune such main SW characteristics as amplitude, phase, wavevector, frequency, etc.
Control of the parameters of SW is possible, for example, when the waves are excited by femtosecond laser pulses \cite{satoh2012directional, Au_directExcitation_PRL2013}.
Since the first experiments with laser excitation of SW, the all-optical pump-probe technique has been used as a powerful toolbox for the detailed investigation of SW, including capture of the wavefront and frequency evolution in the space domain~\cite{kamimaki2017micro, IihamaPRB:2016, filatov2020spectrum}, frequency chirping in the time domain~\cite{Filatov_SWinFe_APL2022}, reconstruction of SW dispersion~\cite{hashimoto_SWaT_NatComm2017, hashimoto2018phaseSWtomographyAPL}.
Furthermore, the shape and size of the excitation laser spot play a critical role in the formation of SW pattern \cite{Jackl_PhysRevX2017, satoh2012directional}.
As an example, an elongated laser pump spot leads to the formation of a quasi-plane wavefront of SW pulse.
This approach unveiled several fundamental processes, inherent to SW and hardly accessible with traditional antennae techniques: bi-reflection of spin waves~\cite{Hioki_Bireflection_CommPhys2020}, coherent oscillation between phonons and magnons~\cite{Hioki_coherent_CommPhys2022}, etc.
On the other hand, the phase of SW could be tuned with double-pump excitation, where the time delay between optical pulses reveals the control other the initial phase of the wave~\cite{Wong_UnidirectionalAPL2014, Kolosvetov_concept_PRApplied2022, Kolosvetov_NORmajority_IEEE2023}, or by using competing excitation mechanisms in single-pump experiments \cite{Yoshimine-JAP2014}.
Thus, single-pulse optical excitation should offer possibilities to generate SW with complex spatial asymmetry patterns of their initial phases, amplitudes, and wavevectors, critical for the design of reconfigurable magnonic logic gates \cite{chumak2022roadmap}.

Here, we present an approach to tune the asymmetry of the initial phase, amplitude and wavevectors of SW at single-pulse optical excitation. 
We use a combination of two SW driving mechanisms, ultrafast demagnetization and ultrafast suppression of magnetic anisotropy, which are inherent to optical excitation of a thin ferromagnetic metallic film.
We analytically show that two types of laser-induced torque maintain different space symmetries, resulting in a non-trivial symmetry of the total torque and parameters of the excited SW.
Using micromagnetic simulations, we further examine, how the spatial asymmetry of initial phase, amplitude, and wavevectors  of the SW vary with the strength of the external magnetic field, its angle with respect to the anisotropy axis, and with the laser spot's diameter.
Moreover, the variation of these parameters reshape the distribution of the SW amplitudes over a range of wavenumbers in certain directions.

\section{\label{sec:theory}Analytical consideration}\label{Sec:Analytics}
\subsection{\label{subsec:geometry}Model of spin waves excitation by a laser pulse}

\begin{figure*}
\includegraphics[width=0.8\linewidth]{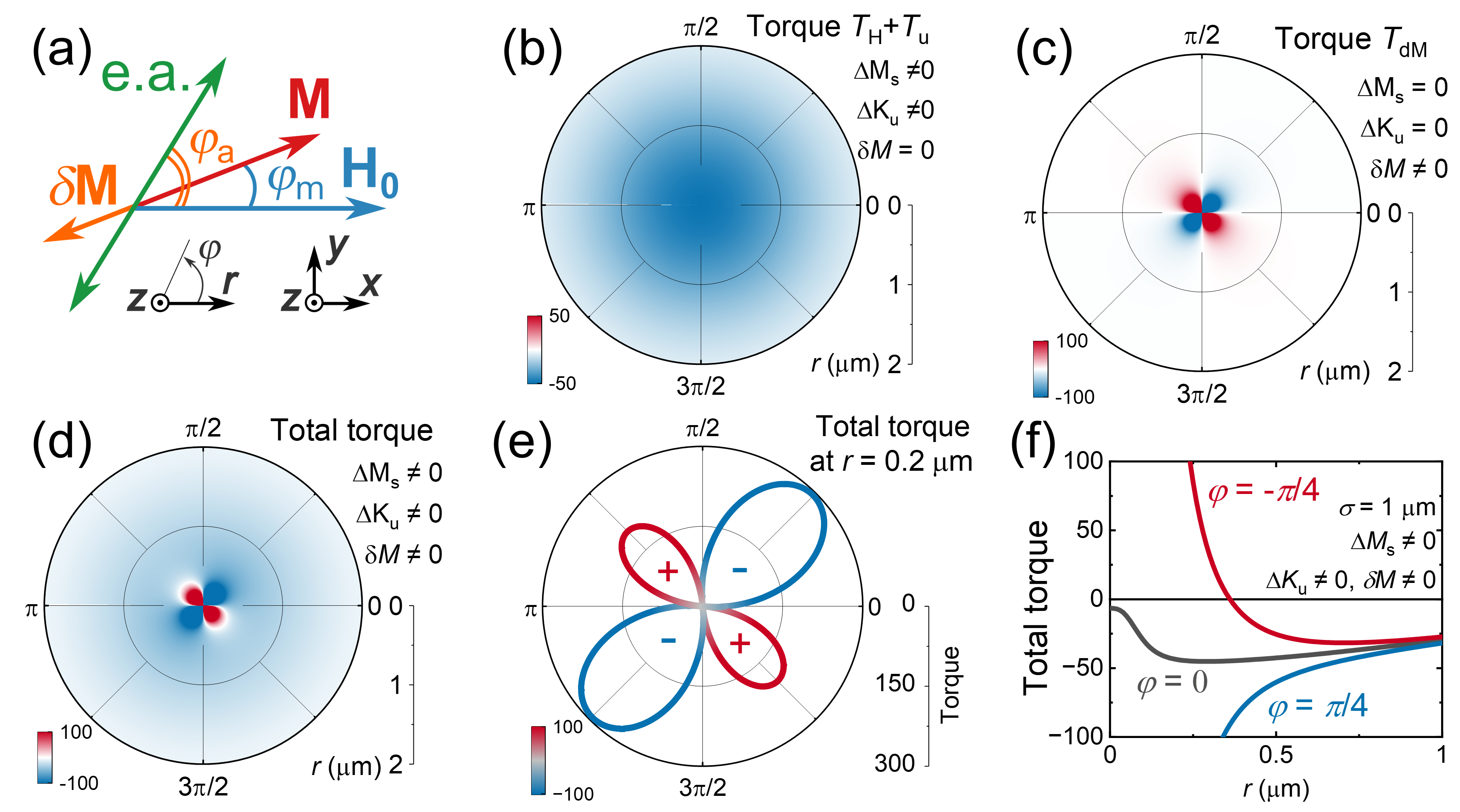}
\caption{\label{fig:geometry}
(a) Geometry of the system.
Arrows: \textbf{H}$_0$ -- external magnetic field, \textbf{M} -- magnetization, e.a. -- easy anisotropy axis, $\delta \textbf{M}$ -- magnetic dipole accounting for ultrafast demagnetization in analytical considerations.
(b) Spatial distribution of the torque $T_H + T_u$, corresponding to the contributions of the Zeeman and anisotropy terms in the total laser-induced torque.
The distribution is obtained analytically with Eq. \eqref{eq:TzTu}.
(c) Spatial distribution of term $T_{dM}$, corresponding to laser-induced demagnetization term in total torque.
The distribution is obtained analytically with Eq. \eqref{eq:T_PID}.
(d) Spatial distribution of total laser-induced torque $T(r, \varphi) = T_H + T_u + T_{dM}$.
(e) Azimuthal distribution of $T(r, \varphi)$ at $r = 0.2\, \mu$m
(f) Radial distribution of $T(r, \varphi)$ at $\varphi = 0, \pm \pi/4$.
On panels (b-f) magnetic field is of magnitude $\mu_0H_0 = 100$ mT, $\varphi_a = \pi /12$, $\sigma = 1 \mu$m. Values of $M_s, K_u, \Delta M_s, \Delta K_u$ used for calculations are listed in Sec.\ref{sec:IIA}; $\delta M = 10^{-17}$\,A/m on (c-f).
}
\end{figure*}

We consider a laterally infinite ferromagnetic thin film with in-plane uniaxial anisotropy (Fig.\ref{fig:geometry}, a).
The saturation magnetization of the material is $M_s$, uniaxial anisotropy parameter is $K_u$.
An external magnetic field $\textbf{H}_0$ is in-plane, and the angle between the easy anisotropy axis and $\textbf{H}_0$ is $\varphi_a$.
The angle between magnetization \textbf{M} and $\textbf{H}_0$ is $\varphi_m$.
Cylindrical coordinate system $(r, \varphi, z)$ is chosen for analytical considerations so that $\textbf{H}_0$ directed along $\varphi = 0$ and $z$ axis is normal to the film surface.
Cartesian coordinate system $(x, y, z)$ is chosen for micromagnetic modeling and complements the cylindrical one with $x$ axis aligned along \textbf{H}$_0$ and $z$ axis being along the film normal.

The impact of the laser pulse is considered as a relative reduction of the magnetic parameters $M_s$ and $K_u$, as it was experimentally observed for metallic thin films ~\cite{Bigot:PRL1996, Carpene_ultrafast_3D_anisotropy_change_PRB_2010, Gerevenkov_PhysRevMaterials2021,shelukhin2020laser}.
In experiments, the initial reduction of the parameters occurs in less than 2 ps~\cite{Bigot:PRL1996, Carpene_ultrafast_3D_anisotropy_change_PRB_2010}.
The process is at least an order of magnitude faster than the typical period of the excited precession in ferromagnets and low-energy spin waves considered in this work.
Thus, we use the instantaneous laser-induced reduction of $M_s$ and $K_u$ in the following consideration.
The maximum values of the reduction are assumed to be $\Delta M_s$ and $\Delta K_u$ for the parameters $M_s$ and $K_u$, respectively.
After the reduction, the magnetic parameters recover to their equilibrium values with a characteristic time of hundreds of picoseconds~\cite{Carpene_ultrafast_3D_anisotropy_change_PRB_2010,Gerevenkov_PhysRevMaterials2021,shelukhin2020laser}.
The time is an order of magnitude longer than the period of the considered low-energy SWs.
Therefore, this recovery process leads to time evolution of SW frequency inside pump spot~\cite{Gerevenkov_PhysRevMaterials2021}, but does not affect the initial asymmetry of phase and amplitude of SW, which is of prime interest here.
Therefore, we used a Heaviside step function $\Theta(t)$ for the temporal laser-induced evolution of $M_s$ and $K_u$ with $t=0$ as the time moment of pulse arrival, which is an appropriate approximation to describe the experimental data~\cite{Au_directExcitation_PRL2013}.
The spatial profiles of $\Delta M_s$ and $\Delta K_u$ are assumed to be round with a Gaussian distribution $G(r)=\exp\left[-0.5(r/\sigma)^{2} \right]$ similar to the typical intensity distribution of the pump in all-optical experiments \cite{khokhlov2019optical,Au_directExcitation_PRL2013,IihamaPRB:2016, Muralidhar_caustics_PRL2021, Khramova_TuningPRB2023}.
Here $\sigma$ characterizes the radius of the pump spot.
The thickness of the film is taken to be less than the penetration depth of light at pump's wavelength to prevent excitation of standing spin waves observed experimentally for thick films \cite{kamimaki2017micro, vanKampen2002PRL}.
Hence, the variation of $\Delta M_s$ and $\Delta K_u$ along $z$ axis can be neglected.
The resulting spatial-temporal profile of magnetic parameters is
\begin{equation}\label{eq:delta_parameters}
    P(r,\varphi,t) = P - \Delta P G(r) \Theta(t) = P(r,t),
\end{equation}
where $P$ stands for $M_s$ or $K_u$.
We omit the dependence on the pump's polarization, as we consider the laser-induced thermal change of magnetic parameters at normal incidence of the pump, which is valid for experiments with metallic thin films~\cite{khokhlov2019optical, KamimakiPRB:2017, Gerevenkov_PhysRevMaterials2021, shelukhin2020laser, shelukhin_MTJ_Nanoscale2022}.

It should be noted that the minimum accessible value of $\sigma$ corresponds to the diffraction limit and is of a hundred nanometers.
Typically, the range of wavenumbers for optically excited SW propagating laterally has the upper limit of $\sqrt{2}/\sigma$ \cite{KamimakiPRB:2017, khokhlov2019optical, chernov2017optical, Jackl_PhysRevX2017}.
Thus, wavenumbers lesser than 10\,rad/$\mu$m are reported in optical experiments, which corresponds to the magnetostatic type of SW.
Nonetheless, exchange perpendicular standing SW are accessible by optical excitation in flat metal films \cite{kamimaki2017micro, vanKampen2002PRL} and complex dielectric structures \cite{chernov2020exchSW_Nanoletters} along with the recently demonstrated optically excited high-speed exchange SW in antiferromagnetic dielectric \cite{hortensius2021coherent}.
Such waves do not propagate laterally in the experiments and are omitted from the consideration below.

\subsection{Azimuthal symmetry of the laser-induced torque}\label{parIB}

The lateral symmetry of the laser-induced SW is defined by the symmetry of initial torque \textbf{T}($r,\varphi$), acting on \textbf{M} at time moment right after the pump excitation $t =0+$~\cite{hashimoto2018phaseSWtomographyAPL, satoh2012directional}.
Non-zero \textbf{T}($r,\varphi$) in the considered system is the result of an abrupt change of the total Gibbs energy $F$ due to pump-induced ultrafast heating and can be found as~\cite{cullity2011introduction}:

\begin{equation}\label{eq:torque_F}
    \textbf{T} = \frac{\partial F}{\partial \varphi_m}.
\end{equation}
Here $F$ contains the terms, corresponding to change of:
(i) Zeeman energy $F_H$ due to ultrafast demagnetization;
(ii) uniaxial magnetic anisotropy energy $F_u$ due to change of anisotropy parameter;
(iii) magnetostatic energy $F_{dM}$ due to spatial distribution of laser-induced demagnetization \cite{hashimoto2018phaseSWtomographyAPL}.
The total free energy is the sum of the above terms: $F = F_H + F_u + F_{dM}$.
Hence, the total torque \textbf{T} is a sum of the corresponding terms $\textbf{T}_H, \textbf{T}_u$, and $\textbf{T}_{dM}$, defined in accordance with Eq. \eqref{eq:torque_F}.

The terms $F_H$ and $F_u$ have the following forms in the considered geometry:

\begin{align}\label{eq:FzFu}
    F_H = - \mu_0 M_s H_0 \cos\varphi_m, \nonumber \\
    F_u = K_u \sin^2(\varphi_a - \varphi_m).
\end{align}
Here, angle $\varphi_m$ depends on $\varphi_a$, as the system is in equilibrium without laser excitation at $t<0$: 
\begin{multline}\label{eq:Tequilibrium}
    \textbf{T}(t<0) = \textbf{T}_H+\textbf{T}_u \\
    = [\mu_0 M_s H_0 \sin\varphi_m - K_u \sin(2\varphi_a - 2\varphi_m)]\hat{\textbf{z}} =  0,
\end{multline}
where $\hat{\textbf{z}}$ is a unit vector along $z$ axis.

The exact analytical expression for $F_{dM}(r,\varphi)$ cannot be obtained in the general case due to its long-range dipole-dipole nature, but could be found numerically with micromagnetic simulations.
To simplify the analytical consideration, we use an approximation from Ref. \cite{hashimoto2018phaseSWtomographyAPL}. 
In particular, we assume that pump-induced demagnetization produces a magnetic dipole $\delta \textbf{M}$ with orientation opposite to \textbf{M} placed at $r=0$ (Fig.~\ref{fig:geometry} a) and having temporal dependence $\Theta(t)$.
Thus, $F_{dM}$ could be represented as Zeeman energy of \textbf{M} in magnetic field $\textbf{H}_{dM}$ produced by the dipole $\delta \textbf{M}$:

\begin{equation}\label{eq:F_PID}
F_{dM} = -\mu_0 \textbf{M} \cdot \textbf{H}_{dM},
\end{equation}
where $\textbf{H}_{dM}$ has a form~\cite{coey2010magnetismBook}:
\begin{equation}\label{eq:Hpid_field}
    \textbf{H}_{dM} = \frac{1}{4\pi} \left[ 3 \frac{(\delta\textbf{M} \cdot \textbf{r})\textbf{r}}{r^5} - \frac{\delta\textbf{M}}{r^3} \right].
\end{equation}

For the following consideration of the spatial distribution $\textbf{T}(r,\varphi)$ we notice that \textbf{H}${_0}$, \textbf{M}, and the effective anisotropy field are aligned in $xy$ plane at a time $t \leq 0$.
Thus, right after the excitation \textbf{T} has non-zero $z$-component only, $i.e.$ $\textbf{T} = T\hat{\textbf{z}}$.
Since the SW initial phase and amplitude distribution is defined by \textbf{T} at $t=0+$, we address directly to its $z$ component $T(r,\varphi)$ below.

Equations (\ref{eq:delta_parameters}-\ref{eq:Tequilibrium}) give the next spatial distribution of torque $T_H + T_u$ at $t=0+$:
\begin{multline}\label{eq:TzTu}
    T_H(r,\varphi) + T_u(r,\varphi) \\ = -\left[ \mu_0 \Delta M_s H_0 \sin \varphi_m - \Delta K_u \sin(2\varphi_a - 2\varphi_m) \right] G(r) \\ = T_H(r) + T_u(r) .
\end{multline}
The equation gives an azimuthally isotropic spatial distribution of torque $T_H+T_u$ with Gaussian dependence on $r$ (Fig.\ref{fig:geometry}b).
Whereas, Eqs.(\ref{eq:torque_F},\ref{eq:F_PID},\ref{eq:Hpid_field}) give:
\begin{multline}\label{eq:T_PID}
    T_{dM}(r,\varphi) = \\ -\frac{3\mu_0}{4\pi r^3} \delta M \left[M_s -\Delta M_s G(r) \right] \sin{(2\varphi - 2\varphi_m)}.
\end{multline}
The distribution of this torque is azimuthally anisotropic with 4-fold symmetry of $\sin{(2\varphi)}$ (Fig.\ref{fig:geometry} c).
In addition, $T_{dM}$ has a radial multiplier of $r^{-3}$ in contrast to the pure Gaussian profile of other torques. 
Furthermore, the magnitude of $\delta M$ depends on the size of demagnetized area, $i.e.$ on radius of the pump $\sigma$.

We note, the equations of $F_{dM}$ and $T_{dM}$ account for the demagnetization twice, in $\Delta M$ and $\delta M$  both. 
It affects the magnitude of the torque $\textbf{T}_{dM}$, but not on its direction and spatial symmetry.
Thus, we discuss the asymmetry of total torque qualitatively in this section.
The quantitative analysis based on micromagnetic calculations is presented in Sec. \ref{Sec:MocroModel}.  

To summarize, the pump-induced torque $T(r,\varphi)$ is a linear combination of azimuthally isotropic and anisotropic terms (\ref{eq:TzTu},\ref{eq:T_PID}).
This means that there can be a combination of the parameters, when the torques \eqref{eq:TzTu} and \eqref{eq:T_PID} cancel each other out in a direction $\varphi_i$, while they add up at $\varphi_i \pm \pi/2$.
It produces a pronounced asymmetry of the total torque (Fig.\ref{fig:geometry} d,e). 
Furthermore, the ratio of the torques (\ref{eq:TzTu},\ref{eq:T_PID}) is defined by the values of $\Delta M_s, \Delta K_u,$ and $\varphi_a$, which, in turn, are tunable with the magnitude and orientation of $\textbf{H}_0$.
We demonstrate the tuning with micromagnetic simulations in Sec. \ref{Sec:MocroModel}B. 
Finally, all this affects the pattern of excited SW, as discussed in Sec. \ref{subsec:SWsymmetyry}.

\subsection{Radial profile of the laser-induced torque}\label{subsec:IC}



In addition to the difference of azimuth dependencies, terms $T_H+T_u$ and $T_{dM}$ possess distinct radial distributions. 
While $T_H+T_u$ [Eq. \eqref{eq:TzTu}] follows the Gaussian distribution, $T_{dM}$ [Eq. \eqref{eq:T_PID}] has an additional factor $r^{-3}$.
In combination with the distinct symmetry of these two contributions, this leads to the existence of an angle $\varphi_i$, where the total torque $T(r, \varphi_i) \neq 0$ and preserves its sign at any $r$, while $T(r, \varphi_i\pm \pi/2) = 0$ at solitary $r=r_i$ and changes sign at this point (Fig.\ref{fig:geometry}f). 
The value of $r_i$ 
is variable with the pump's spot radius $\sigma$, as the latter tunes $G(r)$ in Eqs. \eqref{eq:TzTu} and \eqref{eq:T_PID}, while the factor $r^{-3}$ in Eq.~\eqref{eq:T_PID} remains unchanged.
Therefore, upon excitation with the round laser spot having Gaussian radial distribution, the total induced torque acquires non-Gaussian radial profile (Fig.\ref{fig:geometry}d,e). 
This, in turn, affects the SW pattern in reciprocal space, since the latter is controlled by the spatial distribution of the driving torque \cite{satoh2012directional}.

In the next section, we present results of the micromagnetic modeling to confirm the above analytical predictions about the SW asymmetry tuning with the magnitude and orientation of external field \textbf{H}$_0$ and the pump's spot radius $\sigma$.

\section{Micromagnetic modeling}\label{Sec:MocroModel}

\subsection{Comparison with analytical predictions}\label{sec:IIA}

To illustrate the analytical considerations of the previous section, we performed micromagnetic simulations with the parameters of the material close to those of permalloy at room temperature \cite{zhao2016experimental, Han_DWmotion_bySW_2009_APL}: \mbox{$M_s = 800$ kA/m}, \mbox{$K_u = 1$\,kA/m$^3$}, exchange stiffness \mbox{$A = 1.3 \cdot 10^{-11}$\,J/m}, Gilbert damping parameter $0.01$.
Following experimental observations, the power law
\begin{equation*}
\frac{K_u(\textbf{r},t = 0+)}{K_u(\textbf{r},t<0)} = \left[\frac{M_s(\textbf{r},t = 0+)}{M_s(\textbf{r},t<0)}\right]^a
\end{equation*}
is preserved for metallic films in experiments with femtosecond laser heating~\cite{Carpene_ultrafast_3D_anisotropy_change_PRB_2010, Gerevenkov_PhysRevMaterials2021, shelukhin2020laser}.
Here, we assume the power $a=3$ corresponding to the case of uniaxial anisotropy~\cite{zener1954classical}.
The demagnetization magnitude $\Delta M_s$ is taken to be 10\,\% to be close to typical experimentally observed values, which gives $\Delta K_u= 27.1\%$.
In simulations, we used the thickness of the film of 10\,nm.
A cell size of $5\times5\times 10$\,nm$^3$ is chosen to be smaller than both the magnetostatic exchange length $\sqrt{2A/(\mu _0M_s^2)} =5.7$\,nm and the magnetocrystalline exchange length $\sqrt{A/K_u} = 360$\,nm \cite{abo_definition_Lex_IEEE_2013} in lateral directions.
The infinite size of the film along $x$ and $y$ axes is modeled with 2D periodic boundary conditions.
Micromagnetic simulations are performed using the Object Oriented MicroMagnetic Framework (OOMMF) \cite{OOMMF}.

\begin{figure}
\includegraphics[width=1\linewidth]{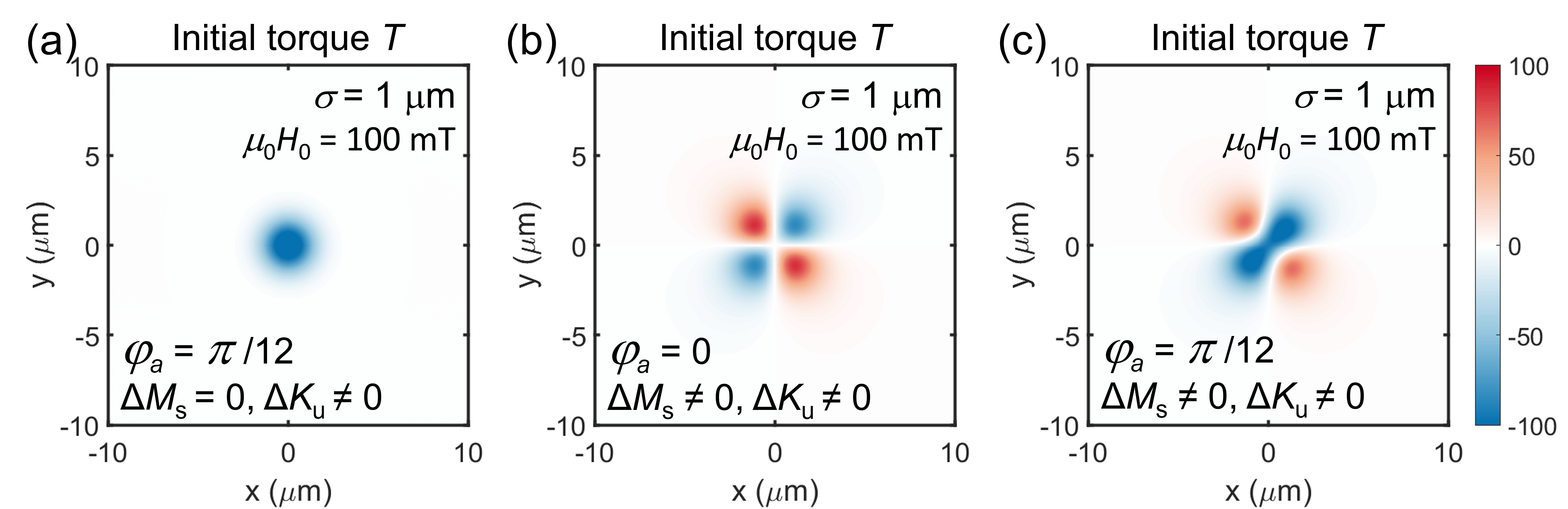}
\caption{\label{fig2:symmetry_variation}
Micromagnetic modeling of spatial distribution of normalized $z$-component of initial torque $T$ at $t = 0+$, $\mu_0 H_0 = 100$\,mT, and $\sigma = 1 \mu$m with next sets of parameters: (a) $\varphi_a = \pi /12, \Delta M_s = 0, \Delta K_u \neq 0$; (b): $\varphi_a = 0, \Delta M_s \neq 0, \Delta K_u \neq 0$; (c): $\varphi_a = \pi /12, \Delta M_s \neq 0, \Delta K_u \neq 0$.
}
\end{figure}

To compare the simulation results with analytical predictions for the symmetry of the term $T_H+T_u$, we set $\varphi_a \neq 0, \pm\pi/2$, and assume $\Delta M_s=0$ to eliminate the contribution of $T_{dM}$.
For simulation of $T_{dM}(x, y)$, we assume $\varphi_a = 0$ to get $T_H+T_u = 0$, as $\varphi_m = 0$ in this case, following the Eq. \eqref{eq:Tequilibrium}.
The symmetries of modeled torques $T_H+T_u$ and $T_{dM}$ fully comply with the analytical model (Fig. \ref{fig2:symmetry_variation}a,b).
Moreover, the total torque $T$ has an asymmetric azimuthal distribution with non-monotonic radial dependence for certain $\varphi$, $e.g.$ $\varphi=-\pi/4$ (Fig.\ref{fig2:symmetry_variation} c), as predicted in Sec. \ref{Sec:Analytics}B,C.

\subsection{Phase and amplitude asymmetry of the laser-induced torque}\label{sec:H_phi_orientation}

\begin{figure}
\includegraphics[width=1\linewidth]{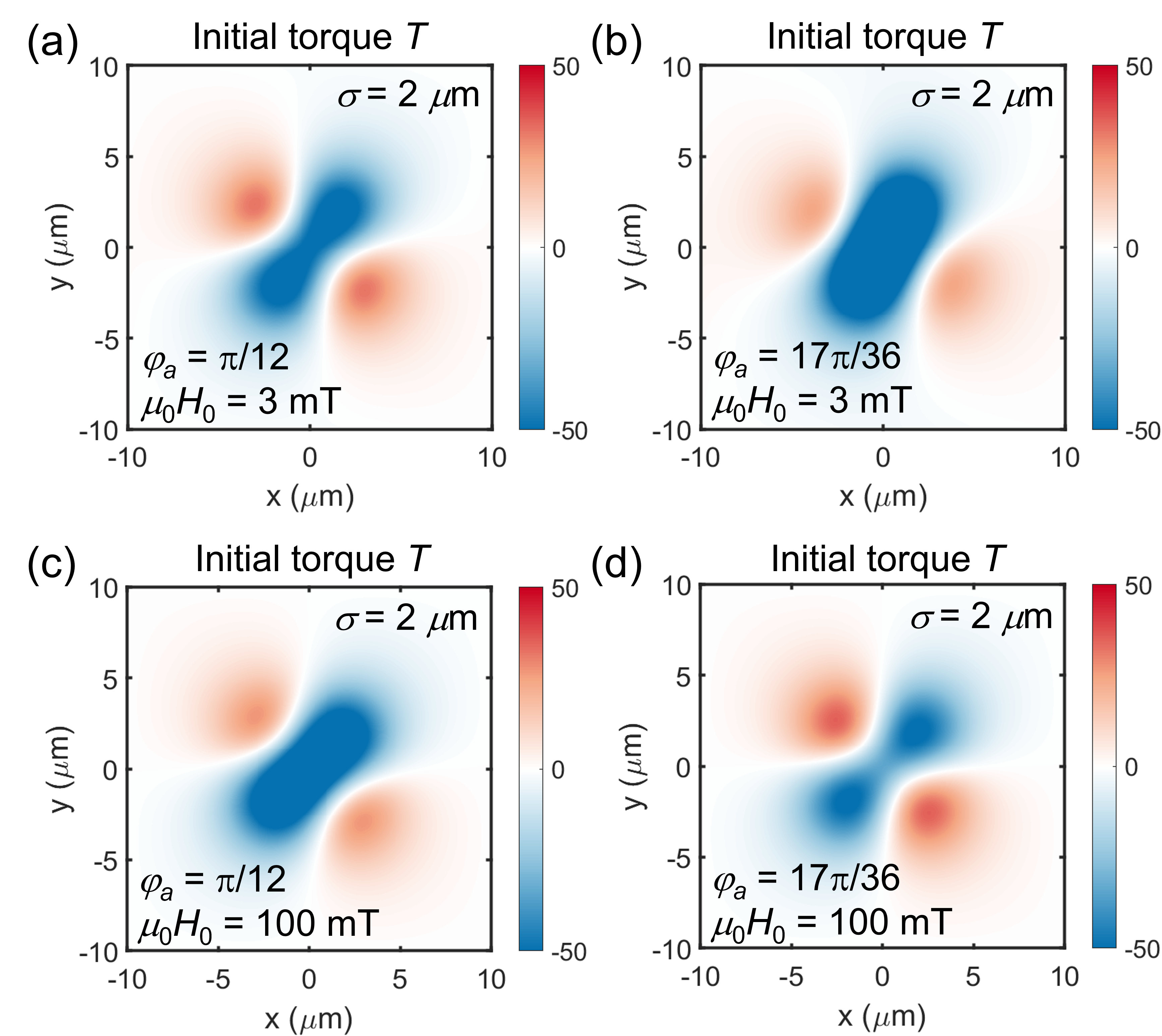}
\caption{\label{fig:vsH_phia}
Micromagnetic modeling of spatial distribution of normalized $z$-component of initial torque $T$ at $t = 0+$, $\sigma = 2\,\mu$m, and various external fields $\textbf{H}_0$:
(a) $\varphi_a = \pi /12$, $\mu_0 H_0 = 3$\,mT; (b) $\varphi_a = 17\pi /36$, $\mu_0 H_0 = 3$\,mT; (c) $\varphi_a = \pi /12$, $\mu_0 H_0 = 100$\,mT; (d) $\varphi_a = 17\pi /36$, $\mu_0 H_0 = 100$\,mT.
}
\end{figure}

As follows from Eqs. \eqref{eq:TzTu} and \eqref{eq:T_PID}, it is possible to tune the space symmetry of $T$ through the magnitude and orientation of the external field ${\bf H}_0$.
In particular, $H_0$ and angle $\varphi_a$ affect the magnitude of torque $T_H + T_u$ \eqref{eq:TzTu}, while the magnitude of $T_{dM}$ \eqref{eq:T_PID} is insensitive to $\varphi_a$.
For example, if \textbf{M} is parallel to \textbf{H}$_0$ at $\varphi_a = 0$, then the isotropic contribution $T_H + T_u$ in \textbf{T} is absent (Fig.\ref{fig2:symmetry_variation} b), as $\varphi_m=0$ in Eqs. (\ref{eq:Tequilibrium},\ref{eq:TzTu}).
Alternatively, $T_H + T_u$ tends to zero in Eq. \eqref{eq:TzTu}, if \textbf{M} is almost aligned with the easy anisotropy axis ($\varphi_m = \varphi_a$) at rather low $H_0$, again resulting in a dominant 4-fold symmetry of the total torque, as shown in Fig.\ref{fig:vsH_phia}a.
A similar situation occurs if $\varphi_a$ is close to $\pi/2$, and $H_0$ is high enough to set $\varphi_m=0$ in Eq. \eqref{eq:Tequilibrium}, $i.e.$ $H_0 > 2K_u/(\mu_0M_s)$ (Fig.\ref{fig:vsH_phia} d).
In turn, the azimuthally isotropic contribution $T_H + T_u$ becomes more prominent at $\varphi_a \approx \pi/2$ in a low field $H_0 < 2K_u/(\mu_0M_s)$ (Fig.\ref{fig:vsH_phia} b) and at intermediate values of $\varphi_a$ even in a high field (Fig.\ref{fig:vsH_phia} c).
Thus, fine control of an external magnetic field magnitude as well as its direction with respect to the anisotropy axis enables a direct manipulation of symmetry of phase and amplitude patterns of optically excited SW in anisotropic thin films.

\subsection{Wavevector asymmetry and impact of laser spot's diameter}\label{sec:spot_diameter}

\begin{figure}
\includegraphics[width=1\linewidth]{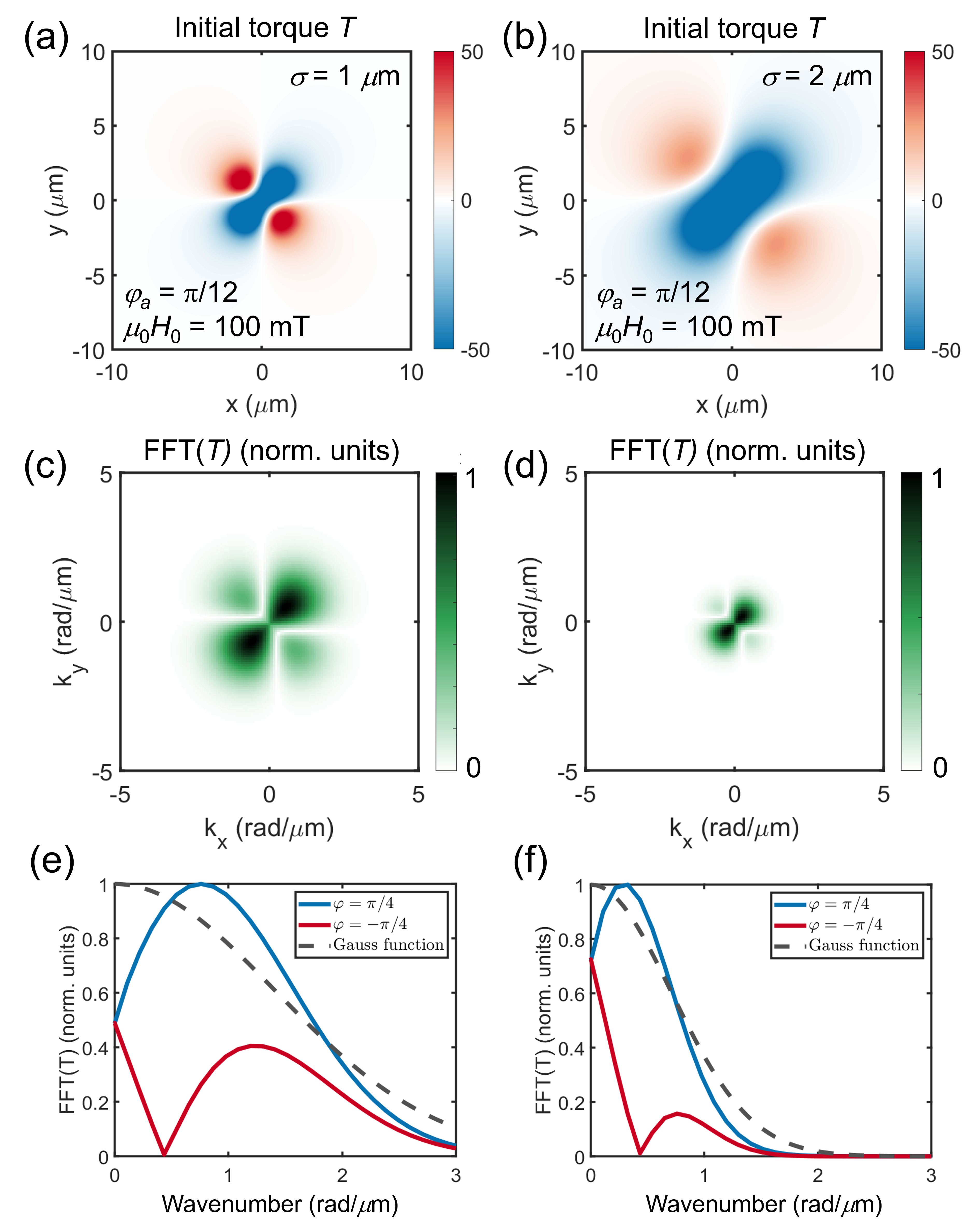}
\caption{\label{fig4:vs_sigma}
(a,b) Micromagnetic modeling of spatial distribution of normalized $z$-component of initial torque $T$ at $t = 0+$, \mbox{$\mu_0H_0 = 100$\,mT}, $\varphi_a = \pi/12$, and $\sigma = 1 \mu$\,m (a) and 2 $\mu$\,m (b).
(c,d) Normalized two-dimensional fast Fourier transforms (FFT) of $T(x,y)$ on panels (a) and (b), respectively.
(e,f) Normalized radial distributions of FFT($T$) at $\varphi = \pm \pi/4$ (blue and red lines) and Fourier transform of Gaussian function $G(r)$ (gray dashed lines) at $\sigma = 1 \mu$\,m (e) and 2 $\mu$\,m (f).
}
\end{figure}

The analytical model in Sec. \ref{subsec:IC} predicts the shaping of the spatial distribution of the driving torque $T$ beyond the Gaussian distribution of the absorbed laser fluence. 
In particular, the pump's radius $\sigma$ defines the ratio between the terms $T_H+T_u$ and $T_{dM}$ in the total initial torque $T(r,\varphi)$ even if the laser spot is round.
Indeed, the variation of $\sigma$ varies not only the spatial distribution of torque $T(x,y)$ (Fig.~\ref{fig4:vs_sigma}a,b), but also its spectrum in reciprocal space $\widetilde{T}(\textbf{k})$ (Fig.~\ref{fig4:vs_sigma}c,d).
As predicted analytically, the distributions demonstrate a solitary radial coordinate $r_i$ in a direction $\varphi_i$ at which the total torque switched its sign ($\varphi_i \approx -\pi/4$ in Fig.~\ref{fig4:vs_sigma}a,b).
This results in the highly non-monotonic distribution of the magnitude of $\widetilde{T}(k)$ along the direction $\varphi_i$ with local spectral density maximum at $k > 0$~(red lines in Fig.~\ref{fig4:vs_sigma}c,f).
At a direction $\varphi_i+\pi/2$ the distribution $\widetilde{T}(k)$ is also non-monotonic, but possess the single maxima at smaller wavenumber $k$~(blue lines in Fig.~\ref{fig4:vs_sigma}c,f).
Thus, driving torque in reciprocal space $\widetilde{T}(\textbf{k})$ appears to be asymmetric.

Moreover, maxima of the spectral density $\widetilde{T}(\textbf{k})$ are shifted to higher $k$ as compared to a Gaussian distribution of the symmetric driving torque (Fig.~\ref{fig4:vs_sigma}e,f).
It means that the combination of torques $T_H+T_u$ and $T_{dM}$ leads to a higher excitation efficiency for SW with higher wavenumbers $k$, in contrast to Gaussian distribution with the maximum excitation efficiency at $k=0$.
Principally, the combination of the parameters could bring $\widetilde{T}(\textbf{k})$ to almost uniaxial pattern in reciprocal space.
It resembles the situation where $\widetilde T(\textbf{k})$ is tunable with different values of $\sigma$ along $x$ and $y$ axes, $i.e.$ by varying the shape of the pump's spot \cite{satoh2012directional} to get a quasi-plane SW pattern in real space \cite{yoshimine2017unidirectional, Hioki_coherent_CommPhys2022}.

\subsection{Asymmetry of the laser-indiced spin waves}\label{subsec:SWsymmetyry}

\begin{figure*}
\includegraphics[width=0.9\linewidth]{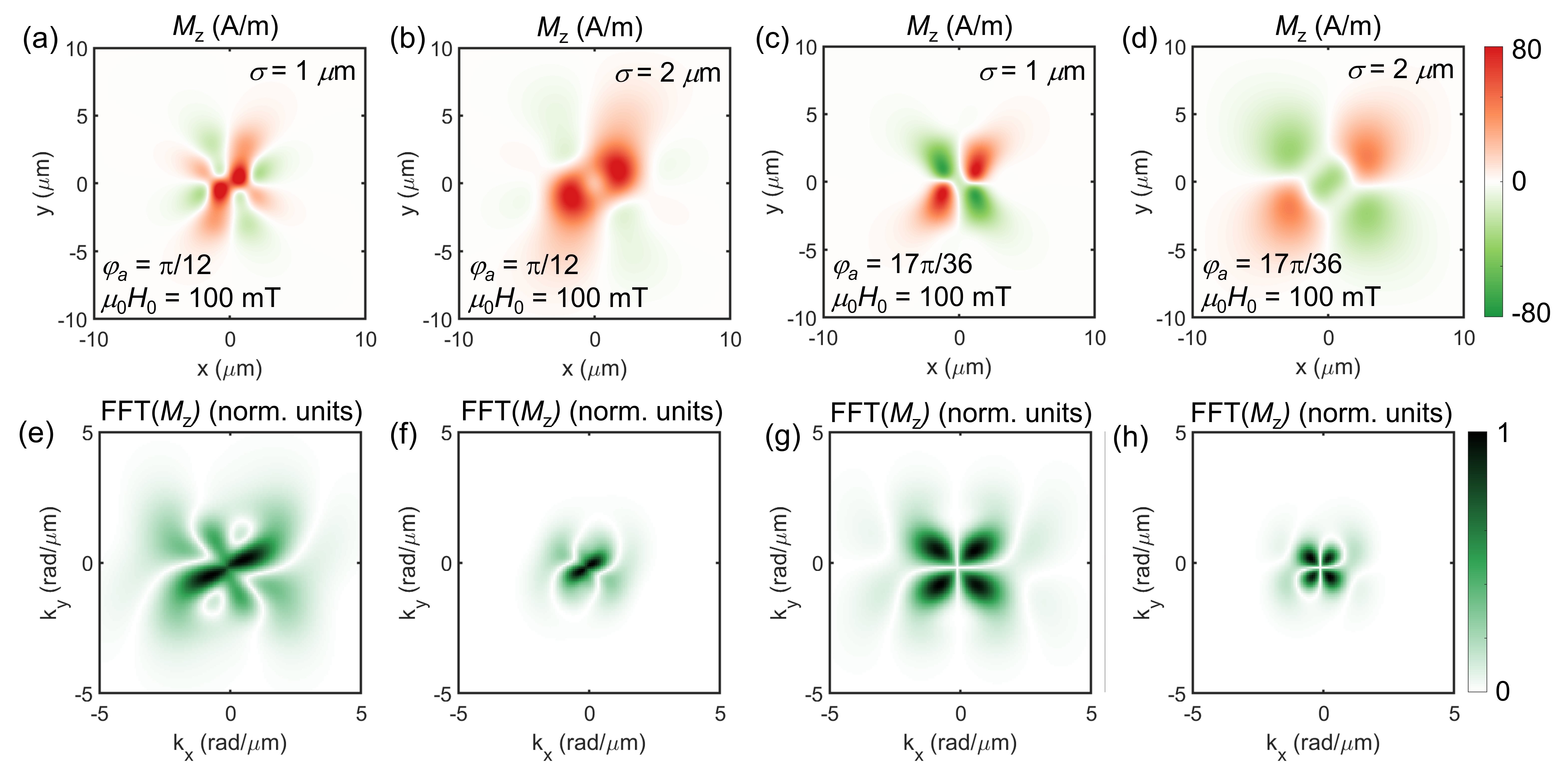}
\caption{\label{fig5:SW_Mxy}
Micromagnetic modeling of SW propagation at $\mu_0H_0 = 100$mT.
(a-d) Spatial distributions of the out-of-plane component of magnetization $M_z(x,y)$ at $t = 1$\,ns at various values of $\varphi_a$ and $\sigma$, shown on panels.
(e-h) Two-dimensional fast Fourier transforms of $M(x,y)_z$ on panels (a-d), respectively.
}
\end{figure*}

Finally, we demonstrate how the spatial symmetry of the spin waves changes upon tuning of the initial torque's symmetry, discussed above.
We consider the spatial distribution of the out-of-plane component $M_z$ and its two-dimensional Fourier transform (Fig. \ref{fig5:SW_Mxy}).
In particular, the intermediate value of $\varphi_a \neq 0, \pm \pi/2$ gives the asymmetric patterns of SW (Fig. \ref{fig5:SW_Mxy} a,b) in correspondence with the discussion in Sec. \ref{sec:H_phi_orientation}.
Moreover, in agreement with the consideration in Sec. \ref{sec:spot_diameter}, the asymmetry of SW pattern varies with the diameter of the pump spot.
The variation is seen clearly in reciprocal space: the wider spot (Fig. \ref{fig5:SW_Mxy} f) excites SW with almost uniaxial distribution of wavevectors with the axis close to $k_x = -k_y$ ($\varphi = -\pi/4$ in real space), while the smaller spot (Fig. \ref{fig5:SW_Mxy} e)  results in a more complex symmetry.
It reproduces the interplay between isotropic and anisotropic terms in \textbf{T} (Sec. \ref{sec:spot_diameter}).
Notably, the quasi-uniaxial $\widetilde{T}(\textbf{k})$ at $\sigma=2\,\mu$m and $\varphi_a = \pi/12$ (Fig. \ref{fig4:vs_sigma}d) reveals the elongated SW pattern in real space (Fig. \ref{fig5:SW_Mxy}b), as predicted.
On the contrary, there is the 4-fold symmetry of SW pattern in real space for various $\sigma$ if $\varphi_a$ is close to $\pi/2$ (Fig. \ref{fig5:SW_Mxy} c,d). 
It is in good agreement with the discussion in Sec. \ref{sec:H_phi_orientation}, as the isotropic contribution $T_H+T_u$ diminishes when the angle and/or magnitude of $\textbf{H}_0$ favors the alignment of \textbf{M} with $\textbf{H}_0$.
Thus, the distribution of $\widetilde{M}_z(\textbf{k})$ in reciprocal space is close to the 4-fold symmetry as well (Fig. \ref{fig5:SW_Mxy} g,h) in consistence with \mbox{$\sin(2\varphi)$-symmetry} of term $T_{dM}$ \eqref{eq:T_PID}. 
The deviation of pattern $\widetilde{M}_z(\textbf{k})$ from pure \mbox{$\sin(2\varphi)$-symmetry}, $i.e.$ with maxima along $\varphi = \pm \pi/4$, is due to anisotropic dispersion of magnetostatic waves.
In particular, the waves possess different signs of group velocity along $x$ and $y$ directions \cite{damon_magnetostatic_1961} with the formation of caustics in intermediate $\varphi \neq \pm \pi/4$ defined by $H_0$.
Notably, the caustics' formation is recently observed also in experiments with a comb of ultrashort laser pulses \cite{Muralidhar_caustics_PRL2021, Khramova_TuningPRB2023}.

\section*{Conclusion}

We show analytically, that the symmetry of laser-excited SW is defined not only by the shape of the optical pump's spot but also by the symmetries of contributions to the excitation torque.
This enables the tuning of SW pattern with orientation and magnitude of the external magnetic field, and the size of the pump's spot without variation of its shape.
The results of analytical considerations are in good agreement with micromagnetic modeling using realistic geometric and magnetic parameters of a thin ferromagnetic metallic film at femtosecond laser fluence.
Hence, the discussed effects could be measured in the experiments with time-resolved scanning magneto-optical Kerr microscopy \cite{Au_directExcitation_PRL2013, khokhlov2019optical, Filatov_SWinFe_APL2022, IihamaPRB:2016, KamimakiPRB:2017, kamimaki2017micro, Shichi_JAP2015}.
In particular, the technique obtains the spatial-temporal distribution of out-of-plane component of magnetization $M_z$ after laser excitation with sub-micrometer and sub-picosecond resolution in space and time.
The distribution $M_z(x,y,t)$ contains information about the initial torque \textbf{T}, discussed here.
From a practical perspective, our results add a degree of freedom in the design of reconfigurable magnonic circuits \cite{grundler2015reconfigurable}.
In particular, we show how the variation of the external magnetic field and laser spot diameter allows us to actively control the initial phase, amplitude, and wavenumbers of SW in different directions. 
As a final remark, we note that the heat-induced change of magnetic anisotropy is a general mechanism inherent to materials of various nature and is not limited to metals only \cite{kalashnikova2021ultrafast, shelukhin2018ultrafast}, which further expands the practical perspectives of the effects considered.
Furthermore, the proposed analytical approach is rather general and allows the consideration of additional contributions that arise from $e.g.$ opto-magnetic effects in nonmetallic media \cite{Kirilyuk_RevModPhys2010, Kalashnikova_UFN2015, Yoshimine-JAP2014}.
Nevertheless, full consideration of stray fields in complex geometries $e.g.$ near the boundaries of waveguides or magnetic defects (domain boundaries, skyrmions, etc.) possible in numerical modeling or with more specific theoretical approaches \cite{vasiliev2007spin, khokhlov2021neel, Gerevenkov_UnidirectionalPRAppl2023, Han_DWmotion_bySW_2009_APL, Hertel_DomainWPRL2004, Lan_SkewPRB2021}.

\section*{ACKNOWLEDGMENTS}

N.E.Kh. and Ia.A.F. acknowledge financial support from the Russian Science Foundation (Project 22-22-00326, \url{https://rscf.ru/en/project/22-22-00326/}).

\section*{COMPETING INTERESTS}
The authors declare no competing interests.

\section*{author contributions}
\textbf{Nikolai E. Khokhlov}: Conceptualization, Methodology, Investigation, Software, Writing- Original draft preparation (equal), Writing- Reviewing and Editing (equal).
\textbf{Iaroslav A. Filatov}: Data curation, Visualization, Writing -- Original draft preparation (equal);
\textbf{Alexandra M. Kalashnikova}: Supervision; Writing -- Reviewing and Editing (equal).

\section*{Data availability statement}
The data that support the findings of this study are available from the corresponding author upon reasonable request.

\bibliography{apssamp}

\end{document}